\documentclass[prc,preprintnumbers,nofootinbib,showpacs,twocolumn]{revtex4}
\usepackage{graphicx}
\usepackage{color}
\usepackage{bm}
\usepackage{amssymb,amsmath}
\usepackage{epstopdf}
\usepackage{epsfig,dcolumn}
\usepackage{array}

\def\sq{\displaystyle{\not} \,q} 
\def\sk{\displaystyle{\not} k} 
\def\sK{\displaystyle{\not} K} 
\def\sP{\displaystyle{\not} P} 
\def\sD{\displaystyle{\not} \Delta } 
\def\beqn{\begin{eqnarray}} 
\def\eeqn{\end{eqnarray}} 
\def\nn{\nonumber} 

\begin{document}
\title{Quark-nucleon dynamics and Deep Virtual Compton Scattering}
\author{M. Gorchtein$^{1}$ and A.P. Szczepaniak$^{1,2}$}
\affiliation{$^1$ Center for the Exploration of Energy and Matter, \\ 
Indiana University, Bloomington, IN 47408\\
$^2$ Department of Physics, 
Indiana University, Bloomington, IN 47405\\} 
\date{\today}

\begin{abstract}
We consider deeply virtual Compton scattering  and deep inelastic scattering in presence of Regge exchanges that are part of the non-perturbative quark-nucleon amplitude.   In particular we discuss contribution from the Pomeron exchange and demonstrate how it leads to Regge scaling of the Compton amplitude.  Comparison with HERA data is given. 
\end{abstract}
\maketitle

\section{Introduction}
\begin{figure}
 \includegraphics[width=2.7in]{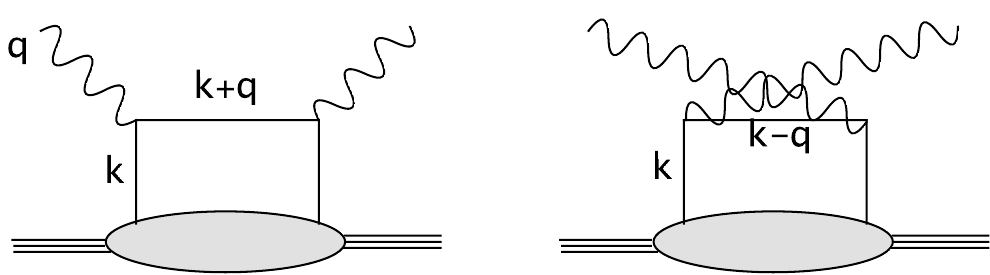}
\caption{Handbag diagram representation of the  Compton amplitude} 
\label{fig:handbag}
\end{figure}

In the past two decades, notable theoretical activity has 
been dedicated to the study of the generalized parton distributions (GPD's) 
\cite{ji1,radyushkin1,goeke,belitsky,diehl,ji2}.
GPD's allow one to access the nucleon structure in a more detailed manner than 
the parton distribution functions (PDF's) studied within DIS paradigm, and are a direct generalization of the latter. To access GPD's, it was proposed to 
study hard exclusive processes like deeply virtual Compton scattering, (DVCS) $e+p\to e+p+\gamma$ \cite{ji3, radyushkin2} or meson electroproduction, 
 $e+p\to e+p+\rho,\omega$,  at high virtuality $Q^2$  of the photon 
originating from the scattered lepton, and low  momentum transfer $t$ between recoiled and target nucleon.  At present DVCS has  been  studied experimentally at HERA \cite{h1,h1_2,zeus,zeus2} 
and Jefferson Lab \cite{clas1,clas2}.
Interpretability of hard exclusive processes in terms of the GPD's that are 
universal objects for all such reaction, is empowered by the collinear 
factorization theorem \cite{collins1, collins2} that, similarly as for DIS, 
allows for a separation of the 
soft hadronic amplitude from perturbative, QCD process with the former  
 leading to four GPD's. To the lowest order in the QCD coupling, $\alpha_s$, the full amplitude then  corresponds to the handbag diagram depicted in Fig.~\ref{fig:handbag}.
  Paratactical applications, however, rest upon, the a priori unknown rate of convergence of the perturbative expansion. At low Bjorken-$x_B$ QCD corrections to the handbag diagram involve  large logarithms  in both $\alpha_s \log Q^2$ and $\alpha_s \log 1/x_B$. While significant progress has been made in devising various  resumation  schemes~\cite{Fadin:1975cb,Balitsky:1978ic,Lipatov:1976zz,Ciafaloni:1987ur,Catani:1989yc,Catani:1989sg,Kwiecinski:1997ee}, to date no first principle solution for the scattering amplitude exists. It is also accepted 
   that the natural physical interpretation of the low-$x_B$ DIS is quite different from that of  parton model description of the  valence region~\cite{dipole1,dipole2,dipole3,kopelyovich,Stasto:2000er,Brodsky:2002ue,Brodsky:2004hi}.  That many orders in the $\alpha_s$ expansion 
    may been needed to describe the low-$x_B$ region is consistent with the ample evidence 
     that  in exclusive electroproduction  nonperturbative phenomena play an important role in the nominally perturbative domain. The structure functions at 
  low-$x_B$ have the behavior characteristic to Pomeron and Regge phenomena, while at fixed momentum transfer,  exclusive photon or meson electroproduction cross 
   sections can be well fitted in terms of simple functions of  $Q^2$ and the center of mass energy,  $W$  rather then $Q^2$ and $x_B$~\cite{donnachie-dosch, Donnachie:2000rz, jenkovszky}.

\begin{figure}
 \includegraphics[width=2.7in]{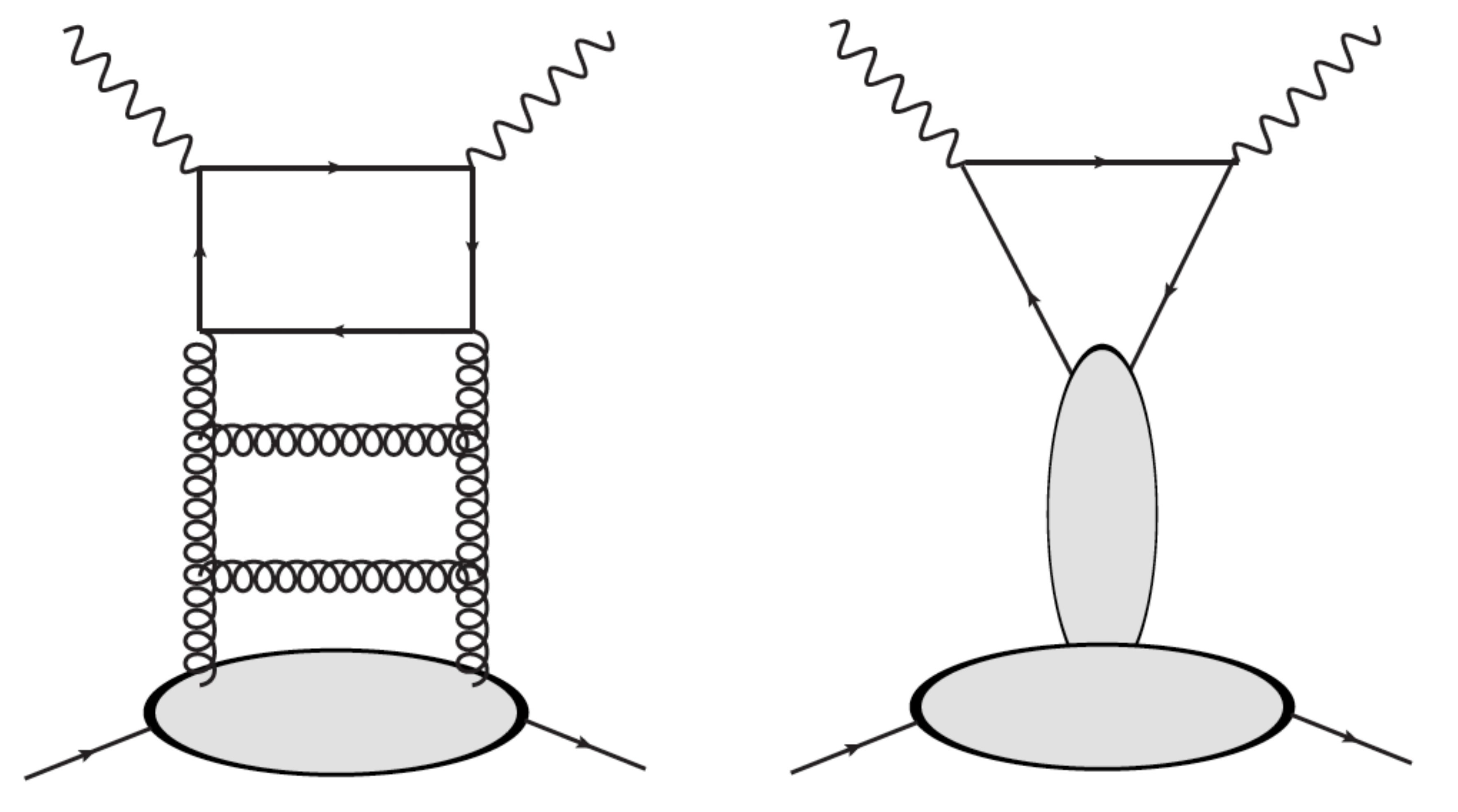}
\caption{Infinite class of perturbative gluon  ladders  (left) is expected to lead to  Regge phenomena and are absorbed into  the quark-nucleon  amplitude (right).  } 
\label{model-1}
\end{figure}

Recently we have proposed a model  in which the diffractive phenomena that are expected 
 to govern the low-$x_B$ DIS are incorporated at the parton nucleon level~\cite{adam-tim, adam}. 
  As discussed above, at the QCD side,  at  low-$x_B$ resumation of gluon ladders leads to complicated evolutions equations. However since  at large center of mass energy, hadronic amplitudes are  known to have a universal Regge scaling, we employ this phenomena to construct an effective parton-nucleon amplitude.  In terms of the QCD description of ~\cite{collins1, collins2},  
 in  the model an infinite class of diagrams, {\it i.e.} those shown on the left panel in 
  Fig.~\ref{model-1} is absorbed into the definition of the parton-nucleon blob and the resulting  electroproduction  amplitude is then  computed from the handbag diagram. The model originates from a study of Regge phenomena at the parton level in the context of DIS~\cite{Landshoff:1970ff,Brodsky:1973hm}.  Such effective parton-nucleon amplitude gives the correct description of low-x structure functions,  surprisingly, however, we have found that in the case of 
  DVCS it breaks collinear factorization, {\it i.e.} Bjorken scaling 
    while it  naturally leads to the Regge-type scaling~\cite{adam-tim, adam}. Upon closer examination, breaking of collinear approximation is not unexpected since it  rests  upon the assumption that parton-nucleon amplitude is a soft function of the invariant parton-nucleon energy, $\hat s$. This is not the case if the amplitude  has Regge-type,  $\hat s^\alpha$, $\alpha > 0$  dependence on $\hat s$.  Such Regge-type scaling of  exclusive amplitudes at large $Q^2$  
    and all $x_B$ as opposed to Bjorken-scaling was in fact predicted by Bjorken and Kogut in ~\cite{Bjorken:1973gc}.   
 
In this paper we focus on applicability of the the model to DVCS in the HERA kinematics, $Q^2/W^2 << 1$.  For description of HERA data on DVCS at low-$x$  two competing 
formalisms are used, Regge models that operate with the soft and hard Pomeron 
 trajectories, as for example in the color dipole and similar 
  models~\cite{dipole1,dipole2,dipole3,kopelyovich,kopelyovich,donnachie-dosch, Donnachie:2000rz, jenkovszky}, and the GPD-based models. To be applied phenomenologically, 
the GPD-based models would include models for  Regge-like background, see {\it e.g.} 
\cite{polyakov-guzey,lech,polyakov-2}.  In general Regge 
background thus represents a systematic effect on the extraction of GPD's. 
Since both kinds of models are more or less successful in 
describing the HERA DVCS data, a question arises on whether the extraction of 
the GPD's is model independent. Moreover, if data allow for interpretation 
without GPD's, as in the model we study or the color dipole models, one 
may question the physical content of all these models. 

The paper is organized as follows. In the following section we discuss the  DVCS amplitude in the handbag approximation and emerging properties of the parton-nucleon amplitude based on Regge phenomenology. Computation of the DIS an DVCS amplitudes is discussed in 
Sec.~\ref{DIS-DVCS} with more details included in the Appendix. Results and comparison with HERA data are presented in Sec.~\ref{results} and followed by summary and conclusions in 
Sec.~\ref{summary}. 
\section{Compton amplitude in the handbag approximation}
The hadronic Compton tensor is given by the matrix element of the time-ordered 
product of two electromagnetic currents,
\begin{eqnarray} 
& & T^{\mu\nu} = i\int d^4ze^{i\frac{q+q'}{2}z}
\langle N|T[J^\nu(z/2)J^\mu(-z/2)]|N\rangle \nonumber \\
\end{eqnarray} 
where $q(q')$ is the four momentum of the incoming (outgoing) photon. We will 
consider both the DIS process that corresponds to the forward virtual Compton 
scattering with both photons spacelike, $q=q'$, $q^2=q'^2\equiv-Q^2<0$, 
and DVCS with $q^2<0,\;q'^2=0$ and $\Delta=q-q'\neq0$. 
The currents are given by 
$J^\mu(z)=\sum_qe_qJ_q^\mu(z),\;J_q^\mu(z)=\bar{\psi}_q(z)\gamma^\mu\psi_q(z)$ 
with $\psi_q$ the quark field operator and $e_q$ the 
quark charge. Using the leading order operator product expansion we 
replace the product of the two currents by the product of two quark field 
operators and a free quark propagator between the photon interaction points 
$z/2$ and $-z/2$, see Fig. \ref{fig:handbag}
In this (handbag) approximation the hadronic Compton amplitude is then given by 
 a convolution 
\beqn
T^{\mu\nu}&=&i\int\frac{d^4K}{(2\pi)^4}
t^{\mu\nu}_{\alpha\beta}(K,q,\delta)
A_{\alpha\beta}(K,\Delta,p,\lambda,\lambda')
\label{compton_qn}
\eeqn
of the quark Compton tensor 
\beqn
&&t^{\mu\nu}_{\alpha\beta}(K,q,\delta)=\label{quark_compton}\\ &&-e_q^2\left[
\frac{\gamma^\nu(\sK+\frac{\sq+\sq'}{2})\gamma^\mu}
{\left(K+\frac{q+q'}{2}\right)^2+i\epsilon}+
\frac{\gamma^\mu(\sK-\frac{\sq+\sq'}{2})\gamma^\nu}
{\left(K-\frac{q+q'}{2}\right)^2+i\epsilon}\right]_{\alpha\beta},
\nn
\eeqn
$\alpha,\beta$ being the Dirac indices, 
and the untruncated, with respect to the parton legs, parton-nucleon amplitude, 
\begin{eqnarray} 
& & A_{\alpha\beta}(K,\Delta,p,\lambda,\lambda') \nonumber \\
& & =-i\int d^4ze^{-iKz}\langle p'\lambda'|T[\bar\psi_\alpha(z/2)\psi_\beta(-z/2)|p\lambda\rangle.
\end{eqnarray} 
 Following \cite{Brodsky:1973hm,adam}, we represent this  amplitude as
\begin{eqnarray} 
& & A_{\alpha\beta}(K,\Delta,p,\lambda,\lambda')
=\int \frac{d\mu^2}{(k'^2-\mu^2+i\epsilon)(k^2-\mu^2+i\epsilon)} \nonumber \\
& & \times \sum_{i}
\left[(\sk\,'+\mu)\Gamma_i^q(\sk\,+\mu)\right]_{\alpha\beta}
\bar u(p')\Gamma_i^Nu(p)
\label{aqN-gammas}
\end{eqnarray} 
where $\Gamma_{i,j}^{q,N}$ are constructed from Dirac  $\gamma$-matrices and the available four-vectors $p,\Delta,k$. The amplitude in Eq.(\ref{aqN-gammas}) gives the correct result in 
perturbation theory, {\it e.g.}  for point-like quark-nucleon interaction. 
For partons bound inside the nucleon, however, $A$ is expected to be suppressed at 
  large-$k^2$ or $k'^2$.  
 This is achieved ~\cite{Brodsky:1973hm,adam}, by  applying to $A$ a generic operator \cite{Brodsky:1973hm} $I_n=(\mu^2)^n\left(\frac{d}{d\mu^2}\right)^n$, 
 so that in Eq.(\ref{aqN-gammas}), 
\beqn
&&\frac{1}{(k'^2-\mu^2+i\epsilon)(k^2-\mu^2+i\epsilon)}\nn\\
&&\to I_n\frac{1}{(k'^2-\mu^2+i\epsilon)(k^2-\mu^2+i\epsilon)}.
\label{eq:inus}
\eeqn
\indent
This method of softening the UV behavior guarantees current
conservation. This would not be the case, for example, if the two propagators were 
 absorbed into a soft quark-nucleon wave function. Furthermore, differentiating the
product of two propagators instead of differentiating each one separately
ensures that the amplitude contains simple poles  that enable to interpolate between the off- and on-shell quark-nucleon amplitudes. 

\subsection{Quark-nucleon amplitude with Regge behavior}
\begin{figure}
\includegraphics[width=2.7in]{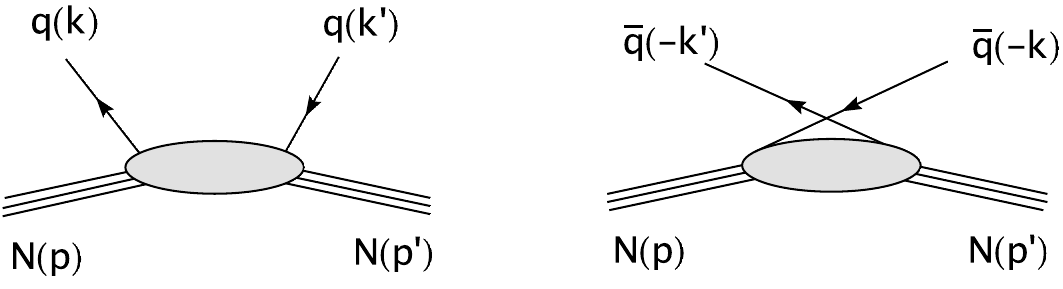}
\caption{\label{qN} Direct and crossed contributions to the 
$quark-nucleon$-scattering amplitude} 
\end{figure}

We proceed by constructing the basis for the scattering 
process $N(p)+q(-k)\to N(p')+q(-k')$ shown in Fig.\ref{qN}.
We account for all possible Dirac-Lorentz structures that can appear in four fermion 
operators.  Furthermore we shall only consider those amplitudes 
 which conserve the quark helicity since helicity-flip amplitudes are suppressed when integrated over in the handbag diagram by a power of $\mu/W$. The structures of interest thus involve 
$\sim\gamma^\mu,\gamma^\mu\gamma_5$ on the quark side only.  Based on $P$, $CP$ and $CPT$ invariance, the quark-nucleon scattering amplitude can be 
decomposed in the basis of six independent tensors each then multiplied by a Lorentz scalar 
function, $a_i$, $i=1,\dots 6$, 
\begin{widetext}
\beqn
A_{qN}&=&\bar{q}\gamma^\alpha q\,\bar{N}\left[a_1\gamma_\alpha
+a_2\frac{i\sigma_{\alpha\beta}\Delta^\beta}{2M}\right] N
+a_3\bar{q}\gamma^\alpha\gamma_5 q\,\bar{N}\gamma_\alpha\gamma_5 N
+\bar{q}\frac{\sK}{M}q\,\bar{N}\left[a_4+a_5\frac{\sK}{M}\right]N
+\bar{q}\frac{\sD\gamma_5}{2M} q\,\bar{N}\frac{\sD\gamma_5}{2M} N a_6, 
\label{basis}
\eeqn
\end{widetext}
were the new four-vectors are defined by, $K=(k+k')/2$, $P=(p+p')/2$, and 
$\Delta=k'-k=p'-p$. The amplitudes $a_i$ are analytic functions of invariants 
$\hat s=(p-k)^2=(P-K)^2$, $\hat u=(p'+k')^2=(P+K)^2$ and $t=\Delta^2$, fixed by the 
condition $\hat s+\hat u+t=2M^2+2\mu^2$ where  $\mu$ is the mass of the effective quarks ({\it c.f.}  Eq.(\ref{aqN-gammas})) and we have explicitly put the quarks on the mass 
shell.  The above basis is equivalent to the form used in \cite{jamarcguichon} for 
elastic electron-proton scattering. In particular the amplitude multiplying $a_3$ is 
chosen to be an  axial vector but can be expressed  in therms 
 of  $\bar{q}\sP q\bar{N}\sK N$ used in \cite{jamarcguichon}. Moreover, 
$\sD\gamma_5/2M$ in front of $a_6$ becomes proportional to $\gamma_5$ for on-shell particles, 
whereas  $\sK/M$ multiplying $a_4$ and $a_5$ reduces to $\mu/M$.

The scalar amplitudes $a_i$ have unitarity  cuts in $\hat s$ and $\hat u$ and at fixed-$t$  
can be represented through a  dispersion representation, 
\beqn
a_i(\hat s,\hat u,t,\mu^2)&=&(2\pi)^4\int ds
\left[\frac{\rho_i^s(s,t,\mu^2)}{s-\hat s-i\epsilon}
+\frac{\rho_i^u(s,t,\mu^2)}{s-\hat u-i\epsilon}
\right]\nn\\
&&+\;{\rm subtractions}
\label{aisu}
\eeqn
with the spectral function  $\rho_i^{s,u}$ being non-zero above some threshold values $s_0(u_0)$ 
in the respective channel. 
Next, we consider the phenomenological consequences of Regge exchanges
on  the asymptotics of the spectral functions at high $s(u)$.   
For fixed $t$, we assume that the on-shell quark-nucleon  helicity
amplitudes follow Regge asymptotics,  ${\it i.e.}$ they are
proportional to $\hat s^{\alpha(t)}$ or $\hat u^{\alpha(t)}$ for large
$\hat s$ or $\hat u$  respectively, $\alpha(t)$ being a Regge trajectory. 
Evaluating the asymptotic behavior of the  amplitudes in  Eq.(\ref{basis}) and comparing with
 the expected behavior in the  Regge limit we find the asymptotic behavior of the spectral functions, 
\beqn
\rho_1^{u,s}&\sim&s^{\alpha_1-1}\nn\\
\rho_2^{u,s}&\sim&s^{\alpha_2}\nn\\
\rho_3^{u,s}&\sim&s^{\alpha_3-1}\nn\\
\rho_4^{u,s}&\sim&s^{\alpha_4}\nn\\
\rho_5^{u,s}&\sim&s^{\alpha_5-1}\nn\\
\rho_6^{u,s}&\sim&s^{\alpha_6}
\label{suasymptotics}
\eeqn
\indent
Note that in the pure collinear kinematics
$\Delta^\mu=(\Delta^+,0,0_\perp)$ (thus for  
$\Delta^2=0$), and for massless quarks and proton, 
the matrix elements at $a_{2,4,6}$ vanish identically. Therefore, they
generally have to be proportional to masses $M,\mu$ or momentum
transfer $\Delta^2$ that is kept constant in Regge limit, and 
the above relations follow. 

An additional constraint on the    
behavior of the spectral functions comes from the Pomeranchuk
theorem which implies  that  asymptotically  $s$ and $u$ channel
amplitudes become equal. The $\hat s-\hat u$ crossing is 
implemented on the level of the quark-nucleon amplitudes according to
\beqn
K&\to&-K\nn\\
\Delta&\to&\Delta\nn\\
\gamma^\alpha&\to&{\cal{C}}\gamma^\alpha{\cal{C}}^\dagger=-\gamma^\alpha\nn\\
\gamma^\alpha\gamma^5&\to&{\cal{C}}\gamma^\alpha\gamma^5{\cal{C}}^\dagger=+
\gamma^\alpha\gamma^5,
\eeqn
with $C$ denoting the charge transformation. For the spectral functions in Eq.(\ref{basis}) Pomeranchuk's theorem then implies,
\beqn
\rho^u_i(s\to\infty)&=& +\rho^s_i(s\to\infty)\;\;\;\;{\rm for}\;\;i=3,4,6,
\nn\\
\rho^u_i(s\to\infty)&=& -\rho^s_i(s\to\infty)\;\;\;\;{\rm for}\;\;i=1,2,5.
\eeqn
We next introduce the $C$-even and  $C$-odd combinations
$\rho_i^\pm\equiv (\rho_i^s\pm\rho_i^u)/2$ which asymptotically behave as, 
\beqn
\rho_1^-\sim s^{\alpha_1-1}&&\rho_1^+\sim s^{\alpha_1-2},\nn\\
\rho_2^-\sim s^{\alpha_2}&&\rho_2^+\sim s^{\alpha_2-1},\nn\\
\rho_3^+\sim s^{\alpha_3-1}&&\rho_3^-\sim s^{\alpha_3-2},\nn\\
\rho_4^+\sim s^{\alpha_4}&&\rho_4^-\sim s^{\alpha_4-1},\nn\\
\rho_5^-\sim s^{\alpha_5-1}&&\rho_5^+\sim s^{\alpha_5-2},\nn\\
\rho_6^+\sim s^{\alpha_6}&&\rho_6^-\sim s^{\alpha_6-1}.
\label{pmasymptotics}
\eeqn
We notice that $\rho_i^-$ and $\rho_i^+$  
correspond to singlet  (valence + sea)  and non-singlet (valence) GPD's, respectively. 
 It is instructive to observe that according to Eq.(\ref{pmasymptotics}), only 
singlet combinations may grow with $s$ in the high energy 
  regime, while the non-singlet ones necessarily vanish at high $s$. 
This fact, trivial in itself since it simply incorporates the symmetry of the 
interaction of the nucleon with highly energetic quark and antiquark, 
has important consequence for collinear factorization.  

In Eq.(\ref{aisu}), convergence of the dispersion integral at high energies 
is governed by asymptotic  energy dependence of
$\rho_i^+/s$ and $\rho_i^-/s^2$.  Combining Eqs.(\ref{pmasymptotics}),(\ref{aisu}), 
 it follows that one can at most expect three subtraction constants,
 for $a_2$, $a_4$ and $a_6$ \cite{footnote}.
The appearance of a finite subtraction constant  that is energy-independent and thus 
has no exponential $t$-dependence would necessarily imply an appearance of 
fixed poles with very mild $t$-dependence in nucleon-nucleon and hadron-nucleon 
scattering. As it was noticed long ago \cite{Brodsky:1973hm,brodsky-subconst}, the 
experimental data do not support  such possibility and we will assume
in the following that these subtraction  constants are zero.

The Pomeron can only contribute to the amplitude $a_1$. 
The amplitude $a_3$ has quantum numbers of an axial vector $a_1$-meson
exchange which has the intercept  $\alpha_{a_1}(0) \approx 0.5$ and
needs no subtraction. The amplitude $a_5$ is crossing-odd and needs no
subtraction.

\section{Regge exchange contribution to DIS and DVCS} 
\label{DIS-DVCS} 
In this section, we will employ handbag formalism and relate the quark-nucleon spectral 
functions $\rho_i^\pm$ to singlet and non-singlet GPD's.
We combine Eqs.(\ref{compton_qn}),(\ref{quark_compton}),(\ref{aqN-gammas}), 
(\ref{aisu}) to obtain the representation for the hadronic Compton amplitude
\begin{eqnarray}
& & T^{\mu\nu}=i\int d\mu^2ds\int d^4K\sum_i \bar u(p')\Gamma_i^Nu(p) \nonumber \\
& & \times \left[\frac{\rho_i^s}{s-(P-K)^2+i\epsilon}
+\frac{\rho_i^u}{s-(P+K)^2+i\epsilon}\right]
\nn\\
& & \times
I_n\frac{Tr\left[(\sK+\frac{\sD}{2}+\mu)t^{\mu\nu}(\sK-\frac{\sD}{2}+\mu)
\Gamma_i^q\right]}
{[(K+\Delta/2)^2-\mu^2+i\epsilon][(K-\Delta/2)^2-\mu^2+i\epsilon]} \nonumber \\
\label{compton_full}
\end{eqnarray} 
Next, we will evaluate the contribution to the hadronic  Compton amplitude
from quark-nucleon amplitude  proportional to $a_1$, {\it i.e.} use $\bar u(p') \Gamma^N_i u(p)  \Gamma^q_i =  
\bar u(p') \gamma_\alpha u(p) \gamma^\alpha$ ($i=1$). 
  This amplitude corresponds to Pomeron (and vector meson)
exchange, so it should give the dominant contribution for DVCS at high 
energies where DVCS data from H1 and ZEUS are available, 
We choose the kinematics~\cite{kinem}  as  $p^\mu=(p^+,0,0_\perp)$ and $q^\mu=(0,Q^2/(2x_Bp^+),Q_\perp)$,  with the usual Bjorken variable $x_B=Q^2/2pq$. 
The trace in Eq.(\ref{compton_full}) can be evaluated using the collinear approximation
\beqn
&&{\rm Tr}(\sk\,'+\mu)t^{\mu\nu}(\sk+\mu)\gamma^\alpha\to
-4g_\perp^{\mu\nu}(k_\perp^2+\mu^2)\frac{Q^2}{2x_BP^+}g^{\alpha+}
\nn\\
&&\times\left[
\frac{1}{(K+\frac{q+q'}{2})^2+i\epsilon}
-\frac{1}{(K-\frac{q+q'}{2})^2+i\epsilon}
\right],\label{trace_a1}
\eeqn
Note that the above trace calculation in collinear kinematics is the same 
for forward (DIS) and non-forward (DVCS) case. Before we proceed, 
we notice that the trace in Eq.(\ref{trace_a1}) is antisymmetric under 
exchanging  $K\to-K$. This implies that only $\rho^-$ spectral density  contributes 
  leading to, 
\begin{widetext}
\beqn
T^{\mu\nu}_{a_1}&=&-4ig_\perp^{\mu\nu}\frac{Q^2}{x_B}
\frac{1}{2P^+}\bar u(p')\gamma^+u(p)
\int d\mu^2ds\int d^4K
I_n\frac{k_\perp^2+\mu^2}
{[(K+\Delta/2)^2-\mu^2+i\epsilon][(K-\Delta/2)^2-\mu^2+i\epsilon]}
\nn\\
&\times&
\rho_1^-(s)
\left[\frac{1}{s-(P-K)^2+i\epsilon} -\frac{1}{s-(P+K)^2+i\epsilon}\right]
\left[\frac{1}{(K+\frac{q+q'}{2})^2+i\epsilon}
-\frac{1}{(K-\frac{q+q'}{2})^2+i\epsilon}\right].
\label{compton_rho-}
\eeqn
\end{widetext}
The fact that the above Compton amplitude depends on the singlet
spectral function $\rho_1^-$ only, is independent of
the collinear approximation: the positive $C$-parity of
the Compton amplitude requires the $C$-even singlet combination 
$\rho_1^-$. On the contrary, the form factor, possessing the odd
$C$-parity only depends on the $C$-odd non-singlet combination $\rho_1^+$.

\subsection{DIS ($\gamma^*p\to\gamma^*p$)}
We next evaluate the amplitude of Eq.(\ref{compton_rho-}) in the forward 
kinematics $\Delta=0$, $q^2=q'^2=-Q^2$.
\begin{eqnarray} 
& & T^{\mu\nu}_{a_1}(\Delta=0)=-4ig_\perp^{\mu\nu}\frac{Q^2}{x_B}
 \frac{1}{2p^+}\bar u(p')\gamma^+u(p) \nonumber \\
& &\times  \int d\mu^2ds \int d^4k
I_n\frac{k_\perp^2+\mu^2}
{(k^2-\mu^2+i\epsilon)^2} \rho_1^-(s) 
\nn\\
& &\times
\left[\frac{1}{s-(p-k)^2+i\epsilon} -\frac{1}{s-(p+k)^2+i\epsilon}\right] \nonumber \\
& & \times \left[\frac{1}{(k+q)^2+i\epsilon}
-\frac{1}{(k-q)^2+i\epsilon}\right].
\label{compton_forward}
\end{eqnarray}
We make the collinear approximation in the hard quark propagators,
\begin{eqnarray} 
& & \frac{1}{(k+q)^2+i\epsilon}\approx\frac{1}{-Q^2+\frac{Q^2}{x_Bp^+}k^++i\epsilon}=
\frac{x_B/Q^2}{\frac{k^+}{p^+}-x_B+i\epsilon}, \nonumber \\
& & \frac{1}{(k-q)^2+i\epsilon}\approx\frac{1}{-Q^2-\frac{Q^2}{x_Bp^+}k^++i\epsilon}=
\frac{-x_B/Q^2}{\frac{k^+}{p^+}x_B-i\epsilon}, \nonumber \\ \label{DIS_c}
\end{eqnarray} 
and obtain (we refer to the Appendix A  for more details),  
\begin{widetext}
\beqn
T^{\mu\nu}(\Delta=0)&=&-4\pi^2g_\perp^{\mu\nu}
\frac{1}{2p^+}\bar u(p')\gamma^+u(p)\Gamma(n)\int_0^1dx(1-x)^{n+1}
\int d\mu^2(\mu^2)^nds\rho_1^-(s,t=0,\mu^2)
\frac{(n+1-x)\mu^2+xs}{[-(1-x)\mu^2-xs]^{n+1}}\nn\\
&\times&
\int dk^+
\left[\frac{1}{\frac{k^+}{p^+}-x_B+i\epsilon}
+\frac{1}{\frac{k^+}{p^+}+x_B-i\epsilon}\right]
\left[\delta(k^+-xp^+)-\delta(k^++xp^+)
\right]
\label{eq:compton_dis}
\eeqn
\end{widetext}
To make a connection to the PDF's, we consider the imaginary part of this 
amplitude. Recalling that the imaginary part of forward Compton tensor 
 proportional to  $-g^{\mu\nu}$ gives 
$\pi W_1\to\pi\frac{1}{2}\sum_qe_q^2[q(x)-\bar q(-x)]$, we identify the parton 
densities with integrals over the  $\rho^-$ or explicitly $s$ and $u$ spectral functions as,
\begin{eqnarray}
& & x_B[q(x_B)-\bar q(-x_B)] =-8\pi^2 \Gamma(n)(-1)^{n+1} (1-x_B)^{n+1}
\nonumber \\
& & \times \int d\mu^2 d\xi (\mu^2)^n
 [\rho_1^s(\frac{\xi}{x_B},0,\mu^2)-\rho_1^u(-\frac{\xi}{x_B},0,\mu^2)] \nonumber \\
& & \times \frac{\xi+(n+1-x_B)\mu^2}{(\xi+(1-x_B)\mu^2)^{n+1}}. 
\end{eqnarray} 
In the above, we changed the integration variable $s$ to $\xi=x_Bs$. Using 
the high energy  asymptotics ({\it c.f.} Eq.(\ref{suasymptotics})) $\rho_1^{s,u}(s) \sim s^{\alpha_P-1}$,
 with $\alpha_P=1+\epsilon$  being the Pomeron trajectory, and pull the $x_B$ 
dependence out of the $\xi$-integral we obtain the experimentally 
 observed asymptotics $F_2(x_B) \sim x_B^{1- \alpha_P} \sim x_B^{-\epsilon}$. This is the result 
 for the singlet PDF. The non-singlet combination will depend on a similar 
integral with the non-singlet spectral function, which at high energy behaves as 
   $\rho^+(s) \sim s^{\alpha_P-2}$, and correspondingly gives 
    $x_B[q(x_B)+\bar q(-x_B)]\sim x_B^{2-\alpha_P} \sim x_B$, as expected. 
Evaluating the real part of the forward Compton amplitude
we obtain the familiar result for DIS,
\beqn
T^{\mu\nu}(\Delta=0)&=&g_\perp^{\mu\nu}
\frac{1}{2p^+}\bar u(p')\gamma^+u(p)\nonumber  \\
&\times&\int_0^1dx\frac{2x}{x^2-x_B^2+i\epsilon}[q(x)-\bar q(-x)]. \nonumber \\ \label{limit}
\eeqn
While the singlet PDF's at low $x$ 
 rise as $x^{-\alpha_P}$,  the singularity at $x \to 0$ is cancelled by one power of $x$ in the numerator of Eq.(\ref{limit})  which makes both the imaginary and real part of the integral  finite~\cite{adam}.
\subsection{DVCS ($\gamma^*p\to\gamma p$): collinear approximation}
Next we evaluate Eq.(\ref{compton_rho-}) in the DVCS kinematics,
$p^\mu=(p^+,0,0_\perp)$, $q^\mu=(0,Q^2/(2x_Bp^+),Q_\perp)$, 
$\Delta^\mu=(-x_Bp^+,0,0_\perp)$, and choose now asymmetric integration 
variable $k$, rather than $K=\frac{k+k'}{2}$,
\begin{widetext}
\beqn
T^{\mu\nu}_{a_1}&=&-4ig_\perp^{\mu\nu}\frac{Q^2}{x_B}
\frac{1}{2p^+}\bar u(p')\gamma^+u(p)
\int d\mu^2ds \int d^4k(k_\perp^2+\mu^2)
I_n\frac{1}
{[k^2-\mu^2+i\epsilon][(k+\Delta)^2-\mu^2+i\epsilon]}
\nn\\
&\times&
\rho_1^-(s)
\left[\frac{1}{s-(p-k)^2+i\epsilon} 
-\frac{1}{s-(p+k+\Delta)^2+i\epsilon}\right]
\left[\frac{1}{(k+q)^2+i\epsilon}
-\frac{1}{(k-q')^2+i\epsilon}\right],
\eeqn
\end{widetext}
Using the collinear approximation for the quark propagator exchanged
between the two photons interaction points we obtain in the case of
DVCS, 
\begin{eqnarray} 
& & \frac{1}{(k+q)^2+i\epsilon}\approx\frac{1}{-Q^2+\frac{Q^2}{x_Bp^+}k^++i\epsilon}
=\frac{x_B/Q^2}{\frac{k^+}{p^+}-x_B+i\epsilon}, \nonumber \\
& & \frac{1}{(k-q')^2+i\epsilon}\approx\frac{1}{-\frac{Q^2}{x_Bp^+}k^++i\epsilon}
=\frac{x_B/Q^2}{-\frac{k^+}{p^+}+i\epsilon}.  \label{DVCS_c} 
\end{eqnarray}
The DVCS amplitude in the collinear approximation is then given by, 
\beqn
T^{\mu\nu}_{a_1}&=&g_\perp^{\mu\nu}
\frac{1}{2P^+}\bar u(p')\gamma^+u(p) \nonumber \\
&\times&\int_0^1dx
\left[\frac{1}{x-x_B+i\epsilon}+\frac{1}{x-i\epsilon}\right]H^+(x,x_B), \nonumber \\ \label{HC} 
\eeqn
and we refer the reader to the Appendix B  for the details of the calculation. 
We identify the singlet GPD $H(x,x_B)$ with,  
\beqn
H^+(x,x_B)&=&(1-x_B/2)\int_0^1dy\int_0^1dz  [q(z)-\bar q(-z)] \nonumber \\
&\times&\delta(x-z-yx_B(1-z)) \label{HH}
\eeqn
which satisfies  the familiar normalization condition, 
\beqn
\int_0^1dxH^+(x,x_B)&=&(1-x_B/2)\int_0^1dx[q(x)-\bar q(-x)]. \nn\\
\eeqn
The factor $(1-x_B/2)$ in the 
definition of the GPD results from the prefactor $1/2P^+$ in the DVCS amplitude. 
Unlike DIS, in the presence of Regge asymptotics, the real part of the 
integral in Eq.(\ref{HC}) is divergent.  This can be seen by first integrating the 
$\delta$-function over $x$, and then performing the integral over $y$. In the limit $z \to 0$ 
the real part of the integral 
\begin{equation} 
 \int_0^1dy\left[\frac{1}{z-x_B+yx_B(1-z)+i\epsilon}
+\frac{1}{z+yx_B(1-z)-i\epsilon}\right] 
\end{equation} 
is  finite, and equal to $\ln(1-x_B)/x_B$ 
Then, given the Regge asymptotics of the PDF,  $[q(z)-\bar q(-z)] \sim z^{-\alpha_P}$ the integral over $z$ diverges. In the case of the DIS amplitude  the quark propagator exchanged between the two photons in  the sum of direct and crossed handbag diagram ({\it c.f.} Fig.~\ref{fig:handbag})  leads to the factor of $x$ in the numerator of Eq.(\ref{limit}). This does not happen in DVCS 
    when  one photon is soft and the sum of the  two  collinear propagators 
    in the DVCS amplitude of Eq.(\ref{HC})  does not vanish when $x \to 0$ and 
     cannot compensate for the rise of the GPD at low $x$. 
     We also note that in the case of the non-singlet GPD, the integral over $x$ instead 
      reduces to  $\sim dx x^{1 - \alpha_P}$ and is therefore convergent. 
 Thus conclude that  for valence GPD's where Regge contributions are suppressed the collinear approximation is adequate and that part of the full  DVCS   amplitude would obey Bjorken scaling. 
As we show in the following section,
  inclusion of Regge contributions  into singlet GPD's leads to Regge scaling. \\

\subsection{DVCS beyond the collinear approximation}
We will use the collinear approximation in the numerator only. 
We combine all four propagators together using Feynman parameters to obtain 
\begin{widetext}
\beqn
&&T^{\mu\nu}=-8ig_\perp^{\mu\nu}\frac{Q^2}{x_B}\frac{1}{2p^+}
\bar u\gamma^+u \int d\mu^2I_n\int ds\rho_1^-(s)
\Gamma(4)\int_0^1dxdydz(1-x)(1-z)^2\\
&&\times\int d^4k
\left[\frac{k_\perp^2+\mu^2}
{([k+zq-(1-z)xp+y(1-x)(1-z)\Delta]^2-z(1-z)Q^2(1-x/x_B-y(1-x))-(1-z)[xs+(1-x)\mu^2])^4}\right.\nn\\
&&\;\;\;\;\;\;\left.-\frac{k_\perp^2+\mu^2}
{([k-zq'-(1-z)xp+y(1-x)(1-z)\Delta]^2-z(1-z)Q^2(x/x_B+y(1-x))-(1-z)[xs+(1-x)\mu^2])^4}\right]\nn
\eeqn
\end{widetext}
We report all the details of the algebra in the Appendix C, and quote  here 
the final result, 
\begin{widetext}
\beqn
T^{\mu\nu}&=&8\pi^2g_\perp^{\mu\nu}\frac{(Q^2)^{\alpha-1}}{x_B^\alpha}
\frac{1}{2p^+}
\bar u\gamma^+u \int d\mu^2\mu^4I_{n-2}\int d\xi\xi^{\alpha-1}
\int_0^{Q^2/x_B}\frac{d\omega}{\omega^{\alpha-1}}\left(1-\frac{x_B}{Q^2}\omega\right)^2
\beta_1^-\left(\frac{Q^2\xi}{x_B\omega}\right) \label{dvcs-f} \\
&\times&\left\{-\frac{\xi+3\mu^2}{[\xi+\mu^2]^3}
\frac{1}{\omega}
\ln\left[\frac{|\xi+\mu^2-\omega(1-x_B)|}{\xi+\mu^2+\omega}
\frac{\xi+\mu^2+Q^2+\omega(1-x_B)}{|\xi+\mu^2+Q^2-\omega|}\right]
\right.\nn\\
&&+\frac{2\mu^2}{[\xi+(1-x)\mu^2]^2}
\left[
\frac{2-x_B}{[\xi+\mu^2-\omega(1-x_B)][\xi+\mu^2+\omega]}
-\frac{2-x_B}{[\xi+\mu^2+Q^2+\omega(1-x_B)][\xi+\mu^2+Q^2-\omega]}
\right]\nn\\
&&\left.+2\frac{\mu^2+Q^2}{\xi+\mu^2}
\left[
\frac{(2-x_B)(\xi+\mu^2+\omega x_B/2)}
{[\xi+\mu^2-\omega(1-x_B)]^2[\xi+\mu^2+\omega]^2}
-\frac{(2-x_B)(\xi+\mu^2+Q^2-\omega x_B/2)}
{[\xi+\mu^2+Q^2+\omega(1-x_B)]^2[\xi+\mu^2+Q^2-\omega]^2}
\right]\right\}\nn,
\eeqn
\end{widetext}
where we changed variables from $s$ to $\xi=xs$, 
from $x$ to $\omega=Q^2 x/x_B$, and factored out the Regge asymptotics 
of the spectral function as $\rho_1^-(s)=s^{\alpha-1}\beta_1^-(s)$ with 
$\beta\to const.$ for $s\to\infty$. 
Analyzing the above formula, we notice that integrals now converge. 
Importantly, large values of $\omega$ do not contribute to the integral because 
of the explicit suppression factor $(1-x_B \omega/Q^2)^2$ and because 
of powers of $\omega$ in the denominator inside the bracket. 
The price to pay for this convergence is the appearance of the explicit scale 
dependence $\sim\mu^2$ in the expressions, as compared to the scale-independent 
results obtained within the collinear approximation. This scale 
dependence is of no surprise since Regge behavior does introduce a scale. 
In the limit $Q^2/\mu^2 >> 1$ it can be shown that the leading  contribution of the Pomeron, $\alpha = \alpha_P$  to this integral is proportional to 
\beqn
T_{DVCS}\sim\frac{1}{Q^2}\left(\frac{Q^2}{x_B}\right)^{\alpha_P} \sim \frac{W^{2\alpha_P}}{Q^2}.
\eeqn
In the following Section, we will confront this parametrization with the DVCS 
data.

\section{Results and Comparison with HERA data}
\label{results} 
The result of the previous section, for $T_{DVCS}$ 
 was obtained in the limit  $Q^2 \to \infty$ at finite $Q^2$ the amplitude is finite but would require 
  knowledge of the spectral decomposition of the quark-nucleon amplitude at finite energies. 
 When comparing to the experimental data at finite $Q^2$ 
 we thus replace  $1/Q^2$ by a   $\sim1/(1+Q^2/Q_0^2)$ 
with some characteristic scale $Q_0^2$ that we will determine from a fit. 
This is in accord with the experimental observation\cite{h1}
\beqn
\sigma_{DVCS}&=&\sigma_0(W^2)^{2\alpha-2}(Q^2)^\delta
\eeqn
with $\delta\approx-1.5$ rather than $-2$.
It is also this form that is used to describe data within phenomenological 
Regge (or color dipole picture-motivated) models 
\cite{donnachie-dosch,jenkovszky}. We will fit the HERA data using the following parametrization 
  for the cross  section
\beqn
\sigma_{\gamma^* p \to \gamma p} =  \sigma_0 \left[
  \left(\frac{W}{W_0}\right)^{\alpha-1} 
\left(\frac{1}{1 + Q^2/Q_0^2} \right) \right]^2
\label{eq:fit_func}
\eeqn
with $W_0 = 20\mbox{ GeV}$. It is worth noting that using the reggized parton-nucleon amplitude 
 in the handbag model we have effectively  "derived" the
 parametrization proposed in  \cite{donnachie-dosch}). 

\begin{figure}
\includegraphics[width=2.5in]{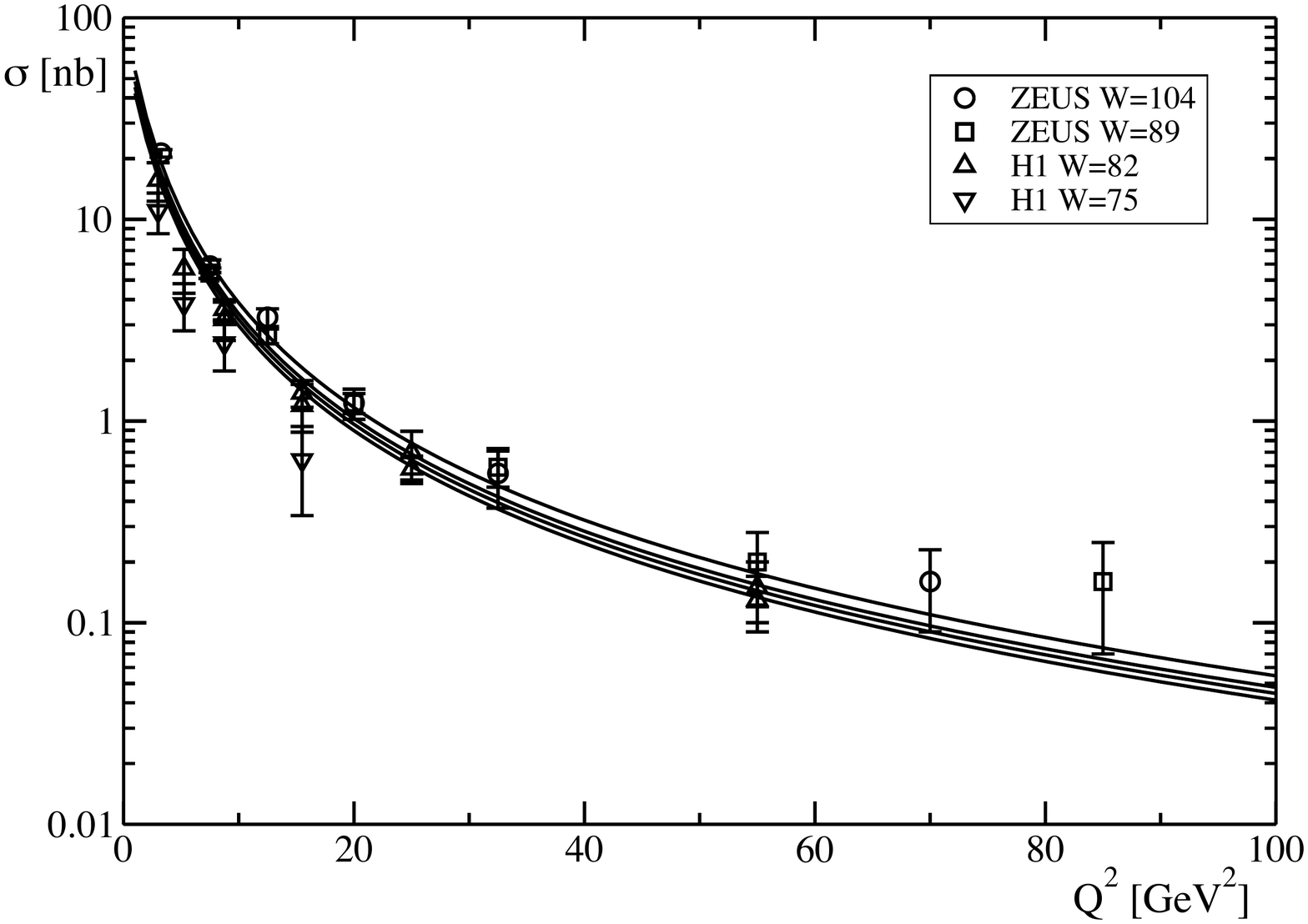}\\
\includegraphics[width=2.5in]{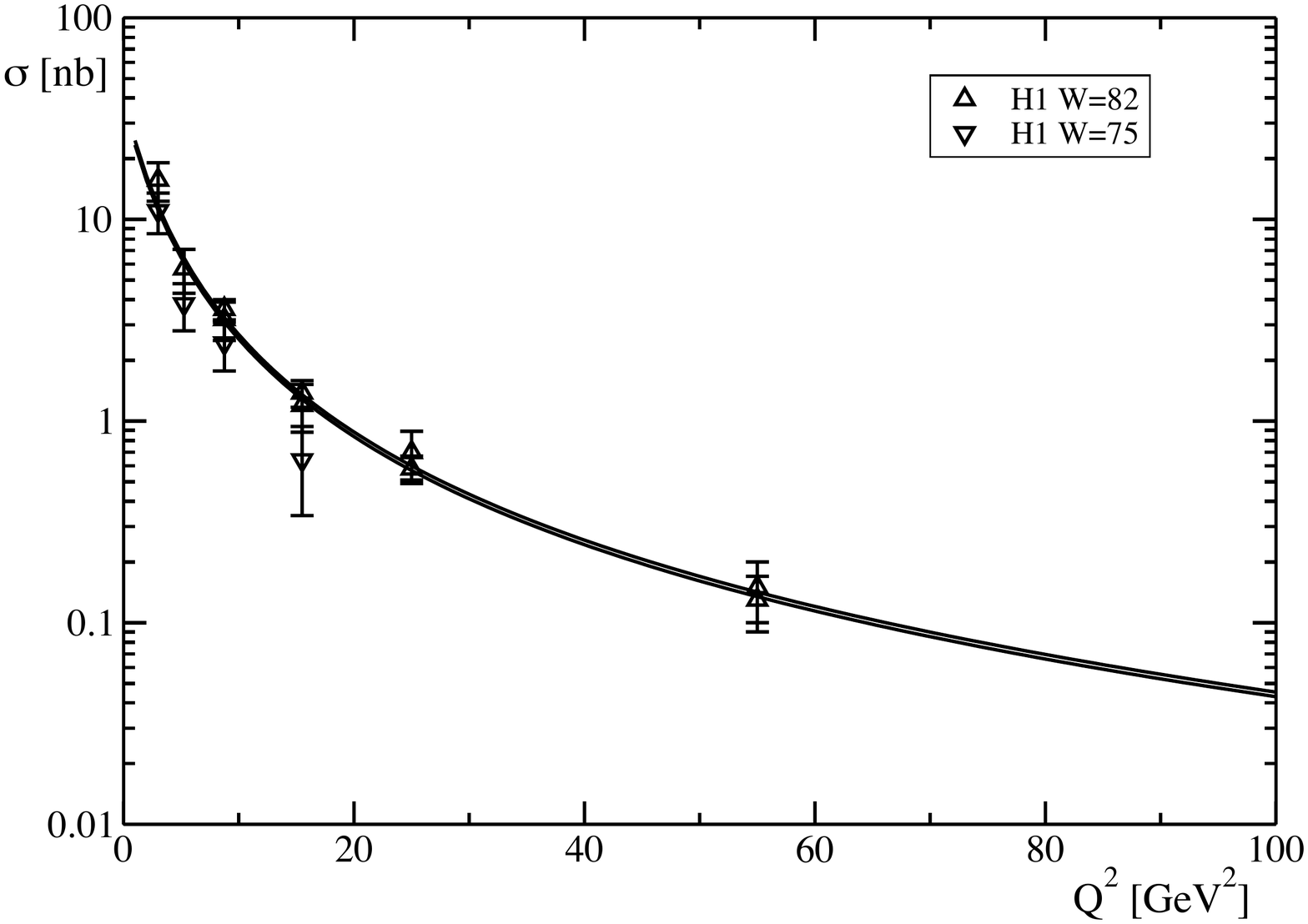}\\
\includegraphics[width=2.5in]{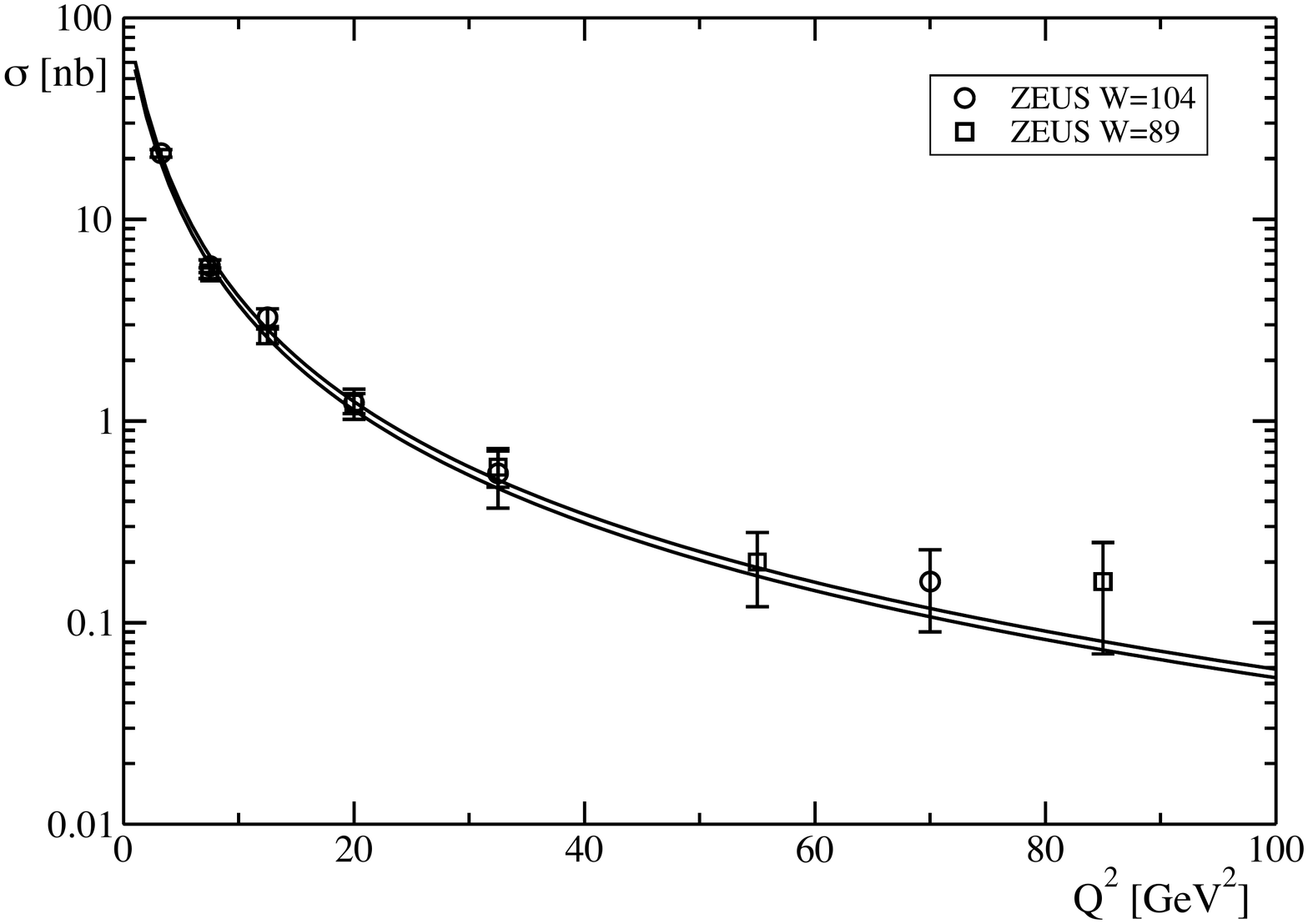}
\label{fig:vs_q2}
\caption{ DVCS cross section as a function of photon virtuality, $Q^2$ for 
various c.m. energies $W$ (in GeV). In the upper panel, we confront the 
combined fit to the H1 and ZEUS data. Solid lines a result of a fit to the 
combined ZEUS and H1 data including both $Q^2$ and $W$ dependence. 
The middle panel displays a similar fit to H1 data alone, whereas the fits to 
ZEUS data alone are shown in the lower panel.} 
\end{figure}

We perform two fits. One is a combined fit to both H1 \cite{h1,h1_2} and 
ZEUS \cite{zeus,zeus2} data.  It gives $\sigma_0 = 28 \pm 4 \mbox{ nb}$,  $Q_0 = 1.51 \pm 0.05 \mbox{GeV}$  and $\alpha-1 = 0.43 \pm 0.03$ and is shown in Figs~\ref{fig:vs_q2}, 
with $\chi^2/d.o.f. = 2.01$. 
\begin{figure}
\includegraphics[width=2.5in]{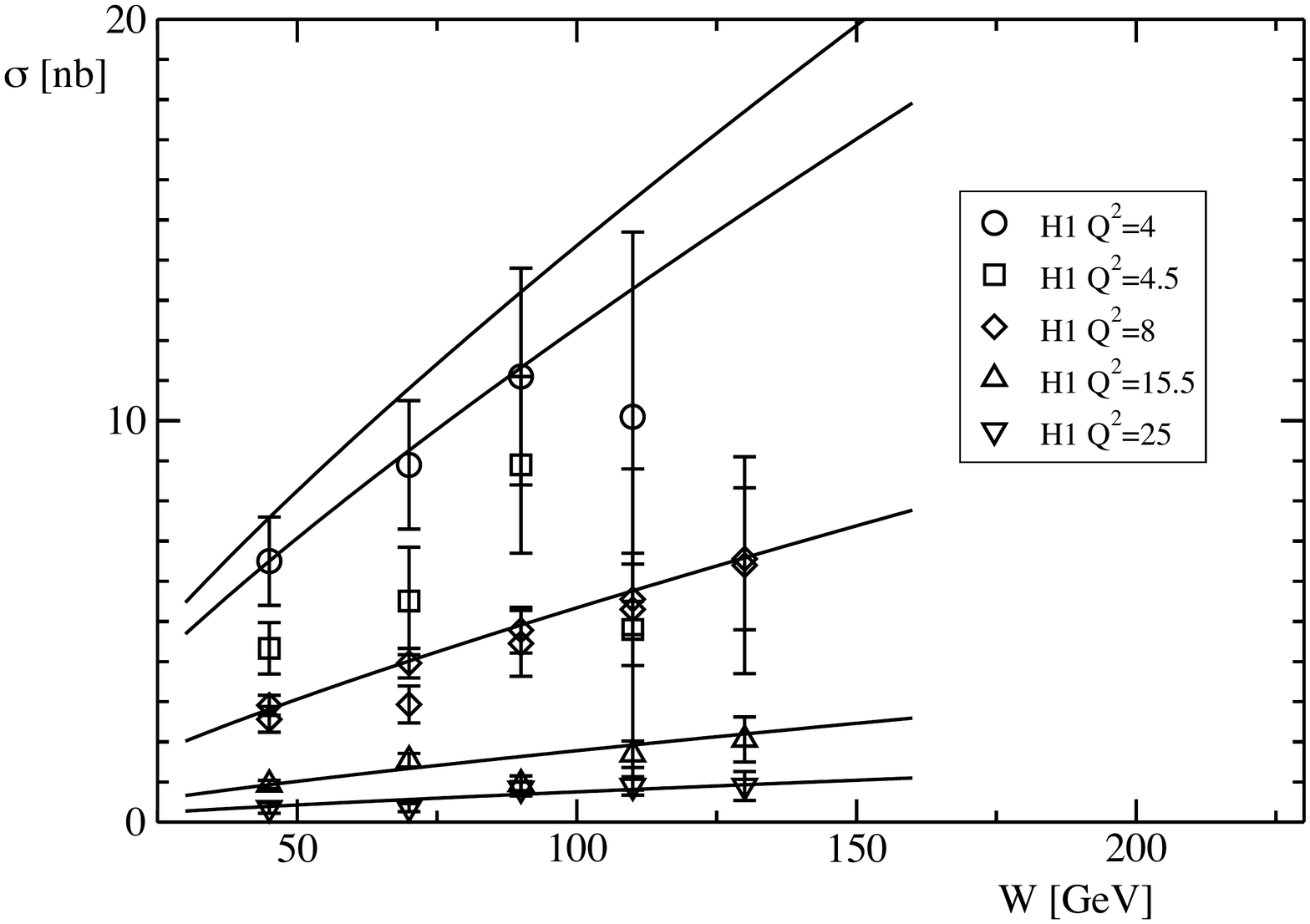}\\
\includegraphics[width=2.5in]{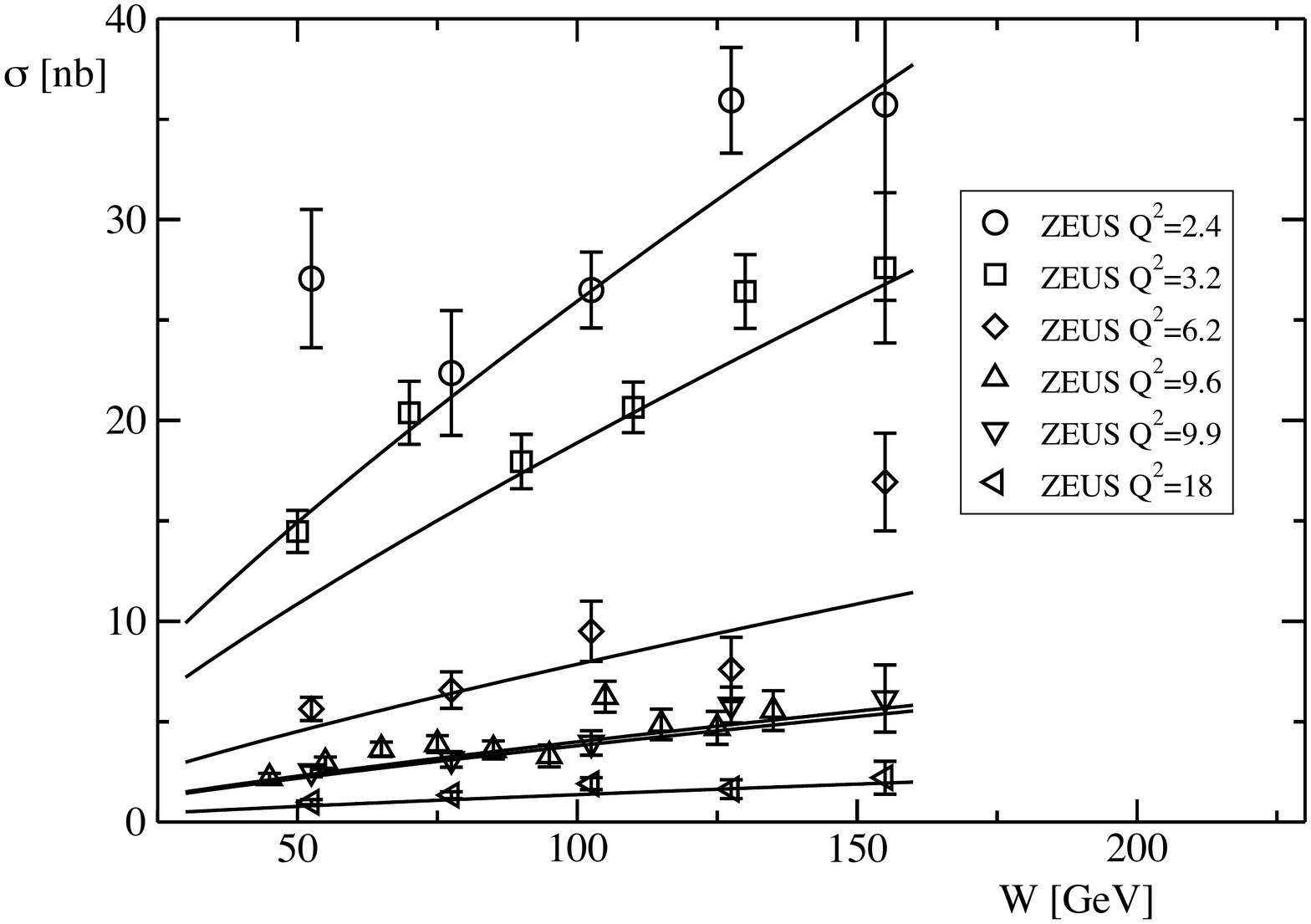}\\
\includegraphics[width=2.5in]{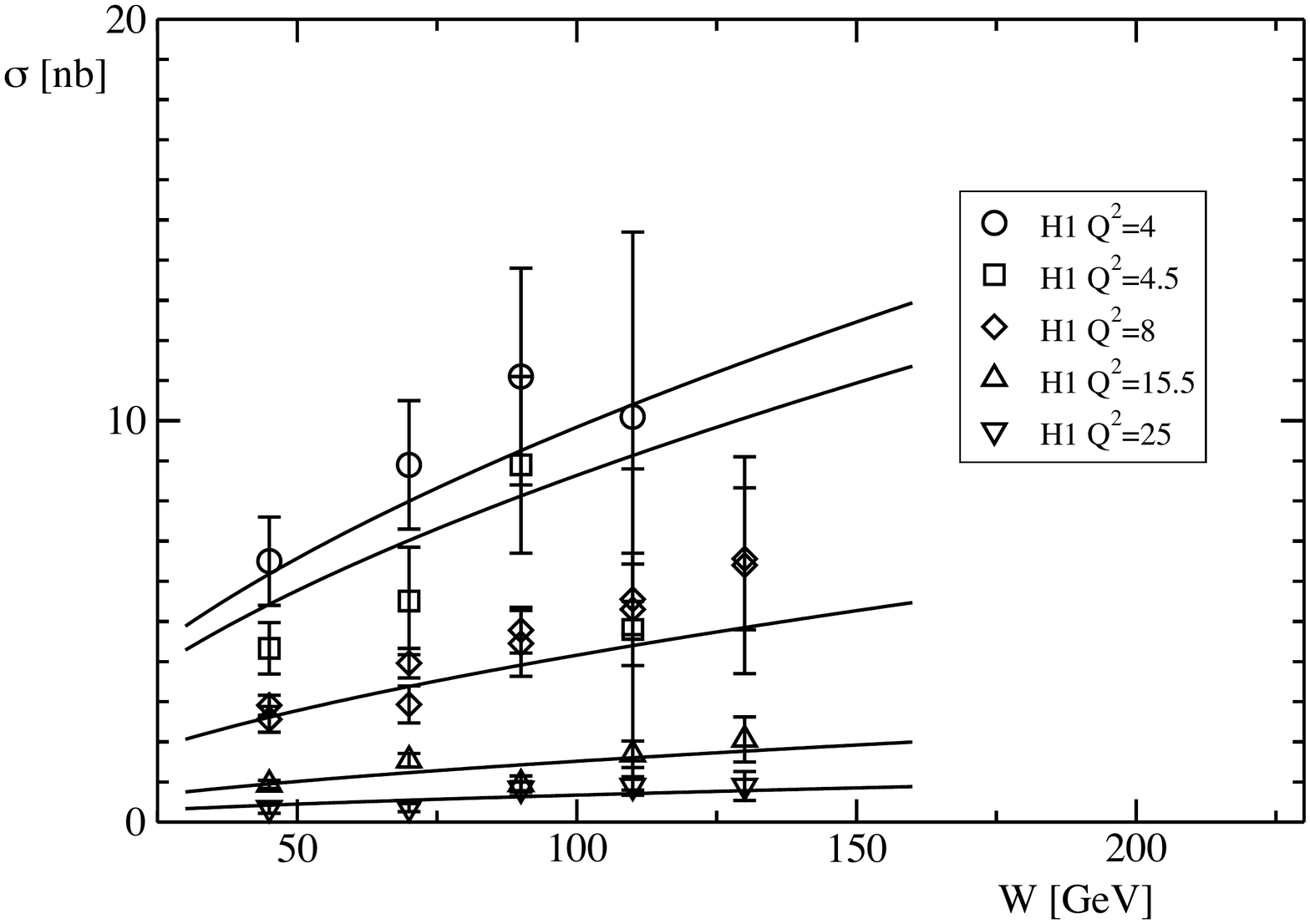}\\
\includegraphics[width=2.5in]{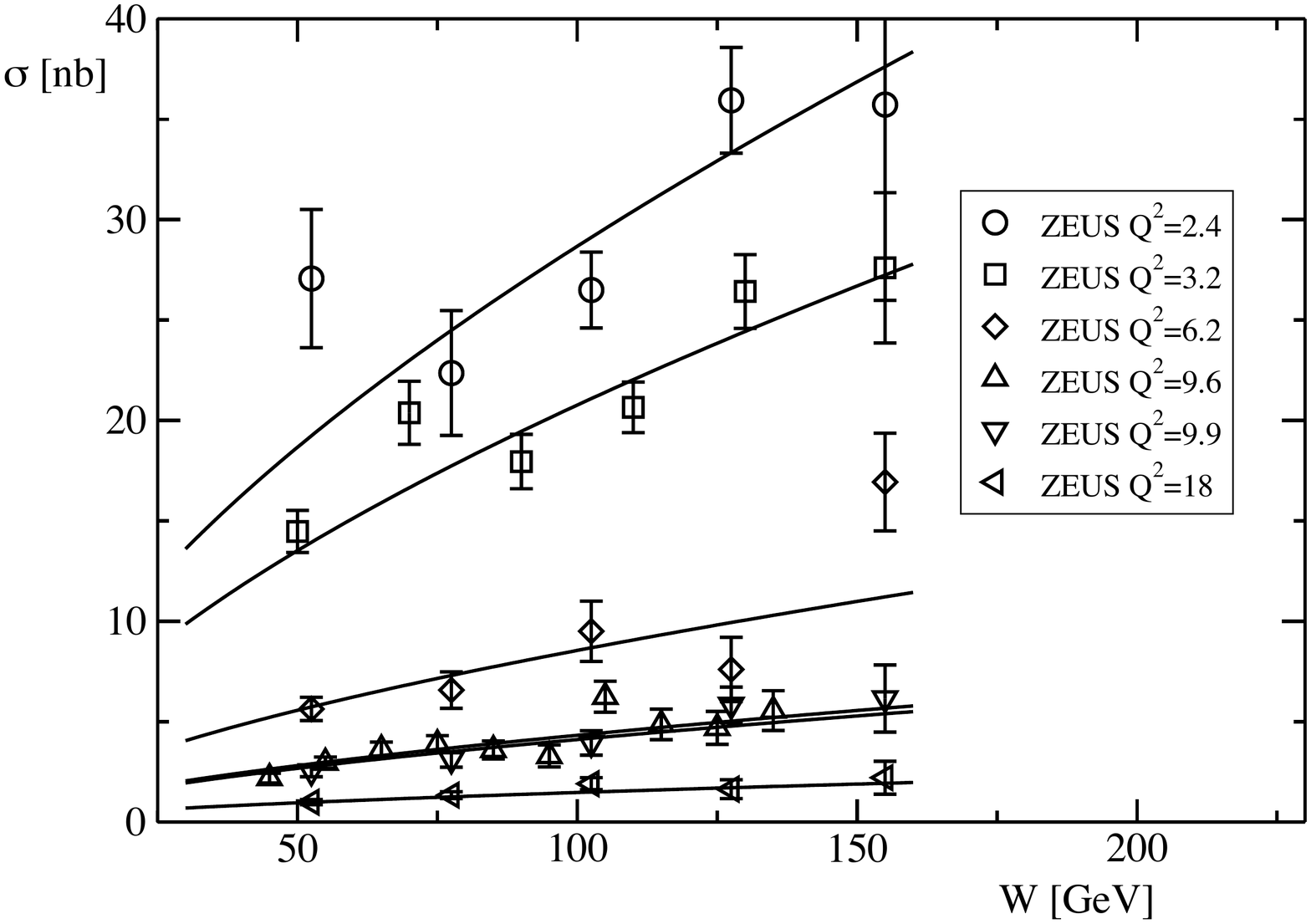}
\label{fig:vs_w}
\caption{$W$-dependence of the DVCS cross section for different values of $Q^2$.
The upper panel displays the comparison of the H1 data to the combined fit to 
both data sets, whereas the second panel from top shows the ZEUS data vs. the 
same fit. The two lower panels confront individual fits to H1 (second lowest 
panel) and ZEUS (lowest panel) to the corresponding data sets.} 
\end{figure} 
The other, is an independent fit to H1 and ZEUS data. For the fit to the H1 data 
alone we obtain  $\sigma_0 = 17 \pm 3 \mbox{ nb}$,  $Q_0 = 1.83 \pm 0.1 \mbox{GeV}$ 
and $\alpha-1 = 0.34 \pm 0.05$ and it is shown in Figs~\ref{fig:vs_q2}, 
\ref{fig:vs_w}, with $\chi^2/d.o.f. = 1.2$. 
For an independent fit to the ZEUS data alone we  find $\sigma_0 = 41 \pm 7 \mbox{ nb}$,  $Q_0 = 1.49 \pm 0.06 \mbox{GeV}$ and $\alpha-1 = 0.34 \pm 0.03$ and it is shown in Figs~\ref{fig:vs_q2}, \ref{fig:vs_w}, with $\chi^2/d.o.f. = 1.1$. 
We observe that both data sets are fitted well with the Regge form of 
Eq.(\ref{eq:fit_func}), as it was found previously in color dipole or Regge 
based studies \cite{donnachie-dosch}. However, the two data sets exhibit different 
normalization (the values of $\sigma_0$). As a result, performing a combined 
analysis we obtain a higher intercept.
\section{Summary} 
\label{summary} 
We presented an analysis of quark-nucleon scattering amplitudes. We
considered a basis of six independent Dirac-Lorentz structures  and  discussed their Regge behavior. 
In particular we have shown that the $C$-odd combinations of the
direct and crossed channels (referred to  as non-singlet combinations) 
follow different Regge asymptotics, as compared to the $C$-even
(singlet) ones. Once embedded into the handbag diagram to describe the 
DVCS amplitude in hard kinematics, we show that  only singlet combinations
contribute, whereas the valence combinations do not appear and require no a priori unknown subtractions. 

We focused on the contribution of a single Pomeron trajectory that dominates at high energies, 
 and have demonstrate that while for DIS the handbag formalism leads
 to the known result, $F_2(x_B)\sim x_B^{-\alpha_P}$, 
 in the case of  DVCS,  the mismatch between quark propagators 
leads to divergent integrals in the collinear approximation. 
If collinear approximation is not used, the model naturally leads to Regge-scaling 
 for DVCS~\cite{Bjorken:1973gc} with 
  $T_{DVCS}\sim Q^{(2\alpha_P-2)}/x_B^{\alpha_P}$, 
  with $\alpha_P=1+\epsilon$ being the Pomeron trajectory. 
Thus we have reproduce the form that  phenomenological Regge models use to 
describe DVCS, and we have illustrated its applicability by fitting the data 
from HERA. 
In he future we plan to extend our phenomenological analysis to
larger values of Bjorken $x_B$, where DVCS was measured at Jefferson
Lab \cite{clas1,clas2}. Since the JLab data is taken at  much lower energies, however, 
 the Pomeron trajectory alone is not expected to be sufficient and other trajectories will  have to be studied. 

\section*{Acknowledgements}

This work was supported in part by the US Department of Energy grant under 
contract DE-FG0287ER40365 and the US National Science Foundation under grant 
PHY-0555232. 

\begin{appendix}
\section{DIS in collinear approximation}
Here we evaluate the forward Compton amplitude of Eq.(\ref{compton_forward}),
\begin{widetext}
\beqn
T^{\mu\nu}_{a_1}(\Delta=0)&=&-4ig_\perp^{\mu\nu}\frac{Q^2}{x_B}
\frac{1}{2p^+}\bar u(p')\gamma^+u(p)
\int d\mu^2ds\int d^4k I_n\frac{k_\perp^2+\mu^2}
{(k^2-\mu^2+i\epsilon)^2}
\nn\\
&\times&
\rho_1^-
\left[\frac{1}{s-(p-k)^2+i\epsilon} -\frac{1}{s-(p+k)^2+i\epsilon}\right]
\left[\frac{1}{(k+q)^2+i\epsilon}
-\frac{1}{(k-q)^2+i\epsilon}\right].
\eeqn
\indent
Using the collinear quark propagators from  Eq.~(\ref{DIS_c})
and introducing  the Feynman parameter $x$, we obtain, 
\beqn
T^{\mu\nu}_{a_1}&=&4ig_\perp^{\mu\nu}
\frac{1}{2p^+}\bar u(p)\gamma^+u(p)
\int d\mu^2ds\rho_1^-(s,\Delta^2,\mu^2)\int dk^+dk^-d^2k_\perp
\left[\frac{1}{\frac{k^+}{p^+}-x_B+i\epsilon}
+\frac{1}{\frac{k^+}{p^+}+x_B-i\epsilon}\right]\nn\\
&\times&
(k_\perp^2+\mu^2)
I_n\frac{1}{[k^2-\mu^2+i\epsilon]^{2}}
\left[\frac{1}{(p-k)^2-s-i\epsilon}
-\frac{1}{(p+k)^2-s-i\epsilon}\right]\nn\\
&=&4ig_\perp^{\mu\nu}
\frac{1}{2P^+}\bar u(p')\gamma^+u(p)\Gamma(n+3)\int_0^1dx(1-x)^{n+1}
\int d\mu^2(\mu^2)^nds\rho_1^-(s,\Delta^2,\mu^2)\nn\\
&\times&
\int dk^+
\left[\frac{1}{\frac{k^+}{p^+}-x_B+i\epsilon}
+\frac{1}{\frac{k^+}{p^+}+x_B-i\epsilon}\right]\nn\\
&\times&
\int dk^-d^2k_\perp
\left[
\frac{(k_\perp^2+\mu^2)}{[(k-yp)^2-ys-(1-y)\mu^2]^{n+3}}
-\frac{(k_\perp^2+\mu^2)}{[(k+yp)^2-ys-(1-y)\mu^2]^{n+3}}
\right].  \label{1} 
\eeqn
\end{widetext}
Finally, Eq.(\ref{eq:compton_dis}) is obtained from Eq.~(\ref{1}) after 
  integrating over  $k^-,k_\perp$ using,
\beqn
\int dk^-d^2k_\perp\frac{1}{(k^2+a^2)^\alpha}
&=&i\pi^2\frac{\Gamma(\alpha-2)}{\Gamma(\alpha)}
\frac{\delta(k^+)}{(a^2)^{\alpha-2}}\nn\\
\int dk^-d^2k_\perp\frac{k_\perp^2}{(k^2+a^2)^\alpha}
&=&-i\pi^2\frac{\Gamma(\alpha-3)}{\Gamma(\alpha)}
\frac{\delta(k^+)}{(a^2)^{\alpha-3}}. \nn\\
\eeqn
The expression in  Eq.(\ref{eq:compton_dis}) follows from Eq.(\ref{1})
after integrating over $k^+$. 
\section{DVCS in collinear approximation}
We evaluate Eq.(\ref{compton_rho-}) in the DVCS kinematics,
$p^\mu=(p^+,0,0_\perp)$, $q^\mu=(0,Q^2/(2x_Bp^+),Q_\perp)$, 
$\Delta^\mu=(-x_Bp^+,0,0_\perp)$, and use $k$ as the integration variable instead of 
 $K=(k+k')/2$,
\begin{widetext}
\beqn
T^{\mu\nu}_{a_1}&=&-4ig_\perp^{\mu\nu}\frac{Q^2}{x_B}
\frac{1}{2p^+}\bar u(p')\gamma^+u(p)
\int d\mu^2ds \int d^4k(k_\perp^2+\mu^2)
I_n\frac{1}
{[k^2-\mu^2+i\epsilon][(k+\Delta)^2-\mu^2+i\epsilon]}
\nn\\
&\times&
\rho_1^-
\left[\frac{1}{s -(p-k)^2+i\epsilon} 
-\frac{1}{s-(p+k+\Delta)^2+i\epsilon}\right]
\left[\frac{1}{(k+q)^2+i\epsilon}
-\frac{1}{(k-q')^2+i\epsilon}\right].
\eeqn
\end{widetext}
We use the collinear approximation of Eq.~(\ref{DVCS_c}) 
 and combine the two quark propagators from the untruncated, quark-nucleon amplitude introducing an integral over a Feynman parameter, 
\begin{eqnarray} 
&& \frac{1}
{[k^2-\mu^2+i\epsilon][(k+\Delta)^2-\mu^2+i\epsilon]}  = \nonumber \\
& & =  \int_0^1dy \frac{1}{[(k+y\Delta)^2-\mu^2+i\epsilon]^{2}},
\end{eqnarray}
 to obtain, 
\begin{widetext}
\beqn
T^{\mu\nu}_{a_1}&=&4ig_\perp^{\mu\nu}
\frac{1}{2p^+}\bar u(p')\gamma^+u(p)\int_0^1dy
\int d\mu^2ds\rho_1^-(s,\Delta^2,\mu^2)\int dk^+dk^-d^2k_\perp
\left[\frac{1}{\frac{k^+}{p^+}-x_B+i\epsilon}
-\frac{1}{-\frac{k^+}{p^+}+i\epsilon}\right]\nn\\
&\times&
(k_\perp^2+\mu^2)
I_n\frac{1}{[(k+y\Delta)^2-\mu^2+i\epsilon]^{2}}
\left[\frac{1}{(p-k)^2-s-i\epsilon}
-\frac{1}{(p+k+\Delta)^2-s-i\epsilon}\right]\nn\\
&=&4ig_\perp^{\mu\nu}
\frac{1}{2p^+}\bar u(p')\gamma^+u(p)\int_0^1dy\Gamma(n+3)\int_0^1dx(1-x)^{n+1}
\int d\mu^2(\mu^2)^nds\rho_1^-(s,\Delta^2,\mu^2)\nn\\
&\times&
\int dk^+
\left[\frac{1}{\frac{k^+}{p^+}-x_B+i\epsilon}
-\frac{1}{-\frac{k^+}{p^+}+i\epsilon}\right]\\
&\times&
\int dk^-d^2k_\perp
\left[
\frac{(k_\perp^2+\mu^2)}{[(k-xp+y(1-x)\Delta)^2-xs-(1-x)\mu^2]^{n+3}}
-\frac{(k_\perp^2+\mu^2)}{[(k+xp'+y(1-x)\Delta)^2-xs-(1-x)\mu^2]^{n+3}}
\right]\nn
\eeqn
\indent
Integrating over $k^-,k_\perp$ results in 
\beqn
T^{\mu\nu}_{a_1}&=&-4\pi^2g_\perp^{\mu\nu}
\frac{1}{2p^+}\bar u(p')\gamma^+u(p)\int_0^1dy\Gamma(n)\int_0^1dx(1-x)^{n+1}
\int d\mu^2(\mu^2)^nds\rho_1^-(s,\Delta^2,\mu^2)
\frac{(n+1-x)\mu^2+xs}{[-(1-x)\mu^2-xs+i\epsilon]^{n+1}}\nn\\
&\times&
\int dk^+
\left[\frac{1}{\frac{k^+}{p^+}-x_B+i\epsilon}
-\frac{1}{-\frac{k^+}{p^+}+i\epsilon}\right]
\left[\delta(k^+-(x+yx_B(1-x))p^+)-\delta(k^++(x(1-x_B)-yx_B(1-x))p^+)
\right]\nn\\
\eeqn
\end{widetext}
The argument of the second $\delta$-function can be brought to the same form 
of the first $\delta$-function by changing integration variables $y\to1-y$ and 
$k^+\to-k^++x_Bp^+$. Finally, the result reads
\begin{widetext}
\beqn
T^{\mu\nu}_{a_1}&=&-4\pi^2g_\perp^{\mu\nu}
\frac{1}{2p^+}\bar u(p')\gamma^+u(p)\int_0^1dy\Gamma(n)\int_0^1dx(1-x)^{n+1}
\int d\mu^2(\mu^2)^nds\rho_1^-(s,\Delta^2,\mu^2)
\frac{(n+1-x)\mu^2+xs}{[-(1-x)\mu^2-xs+i\epsilon]^{n+1}}\nn\\
&\times&
2\left[\frac{1}{x-x_B+yx_B(1-x)+i\epsilon}
+\frac{1}{x+yx_B(1-y)-i\epsilon}\right]
\eeqn
\end{widetext}
which corresponds to Eq. (\ref{HC}) with $H$ defined in Eq.(\ref{HH}). 
\section{DVCS beyond the collinear approximation}
We use the collinear approximation in numerator of Eq.(\ref{compton_full}) and 
 combine all four propagators using Feynman parameters, 
 \begin{widetext}
\beqn
&&T^{\mu\nu}=-8ig_\perp^{\mu\nu}\frac{Q^2}{x_B}\frac{1}{2p^+}
\bar u\gamma^+u \int d\mu^2I_n\int ds\rho_1^-(s)
\Gamma(4)\int_0^1dxdydz(1-x)(1-z)^2\\
&&\times\int d^4k
\left[\frac{k_\perp^2+\mu^2}
{([k+zq-(1-z)xp+y(1-x)(1-z)\Delta]^2-z(1-z)Q^2(1-x/x_B-y(1-x))-(1-z)[xs+(1-x)\mu^2])^4}\right.\nn\\
&&\;\;\;\;\;\;\left.-\frac{k_\perp^2+\mu^2}
{([k-zq'-(1-z)xp+y(1-x)(1-z)\Delta]^2-z(1-z)Q^2(x/x_B+y(1-x))-(1-z)[xs+(1-x)\mu^2])^4}\right]
\nn\\ &&\nn
\eeqn

Integration over $d^4k$ results in
\beqn
T^{\mu\nu}&=&8\pi^2g_\perp^{\mu\nu}\frac{Q^2}{x_B}\frac{1}{2p^+}
\bar u\gamma^+u \int d\mu^2\mu^4I_{n-2}\int ds\rho_1^-(s)
\Gamma(3)\int_0^1dxdydz(1-x)^3\\
&\times&\left\{
\frac{1-z}{[xs+(1-x)\mu^2+zQ^2(1-x/x_B-y(1-x))]^3}
-\frac{1-z}{[xs+(1-x)\mu^2+zQ^2(x/x_B+y(1-x))]^3}\right.\nn\\
&&\left.+
\frac{3(\mu^2+z^2Q^2)}{[xs+(1-x)\mu^2+zQ^2(1-x/x_B-y(1-x))]^3}
-\frac{3(\mu^2+z^2Q^2)}{[xs+(1-x)\mu^2+zQ^2(x/x_B+y(1-x))]^3}\right\}\nn\\ &&\nn
\eeqn

Next the $y$ integral can be done to obtain, 
\beqn
T^{\mu\nu}&=&8\pi^2g_\perp^{\mu\nu}\frac{1}{x_B}\frac{1}{2p^+}
\bar u\gamma^+u \int d\mu^2\mu^4I_{n-2}\int ds\rho_1^-(s)
\Gamma(3)\int_0^1dx(1-x)^2\frac{dz}{z}\\
&\times&\left\{(1-z)\left[
\frac{1}{[xs+(1-x)\mu^2+zQ^2(x-x/x_B)]^2}
-\frac{1}{[xs+(1-x)\mu^2+zQ^2(1-x/x_B)]^2}\right.\right.\nn\\
&&\left.+\frac{1}{[xs+(1-x)\mu^2+zQ^2(1-x+x/x_B)]^2}
-\frac{1}{[xs+(1-x)\mu^2+zQ^2x/x_B]^2}\right]\nn\\
&&+2(\mu^2+z^2Q^2)\left[
\frac{1}{[xs+(1-x)\mu^2+zQ^2(x-x/x_B)]^3}
-\frac{1}{[xs+(1-x)\mu^2+zQ^2(1-x/x_B)]^3}\right.\nn\\
&&\left.\left.+\frac{1}{[xs+(1-x)\mu^2+zQ^2(1-x+x/x_B)]^3}
-\frac{1}{[xs+(1-x)\mu^2+zQ^2x/x_B]^3}\right]\right\}\nn
\eeqn
\indent
and finally $z$ integral yields, 
\beqn
T^{\mu\nu}&=&8\pi^2g_\perp^{\mu\nu}\frac{1}{x_B}\frac{1}{2p^+}
\bar u\gamma^+u \int d\mu^2\mu^4I_{n-2}\int d\xi\xi^{\alpha-1}
\beta_1^-(\frac{\xi}{x})
\int_0^1\frac{dx}{x^\alpha}(1-x)^2\\
&\times&\left\{-\frac{\xi+(3-x)\mu^2}{[\xi+(1-x)\mu^2]^3}
\ln\left[\frac{\xi+(1-x)\mu^2+Q^2(x-x/x_B)}{\xi+(1-x)\mu^2+Q^2x/x_B}
\frac{\xi+(1-x)\mu^2+zQ^2(1-x+x/x_B)}{\xi+(1-x)\mu^2+zQ^2(1-x/x_B)}\right]
\right.\nn\\
&&+\frac{2\mu^2}{[\xi+(1-x)\mu^2]^2}
\left[
\frac{1}{\xi+(1-x)\mu^2+Q^2(x-x/x_B)}
-\frac{1}{\xi+(1-x)\mu^2+Q^2(1-x/x_B)}\right.\nn\\
&&\left.+\frac{1}{\xi+(1-x)\mu^2+Q^2(1-x+x/x_B)}
-\frac{1}{\xi+(1-x)\mu^2+Q^2x/x_B}\right]\nn\\
&&+\frac{\mu^2+Q^2}{\xi+(1-x)\mu^2}
\left[
\frac{1}{[\xi+(1-x)\mu^2+Q^2(x-x/x_B)]^2}
-\frac{1}{[\xi+(1-x)\mu^2+Q^2(1-x/x_B)]^2}\right.\nn\\
&&\left.\left.+\frac{1}{[\xi+(1-x)\mu^2+Q^2(1-x+x/x_B)]^2}
-\frac{1}{[\xi+(1-x)\mu^2+Q^2x/x_B]^2}\right]\right\}\nn,
\eeqn
\end{widetext}
where we changed variables from $s$ to $\xi=xs$ and factored out the Regge asymptotics 
of the spectral function as $\rho_1^-(s)=s^{\alpha-1}\beta_1^-(s)$ with 
$\beta\to const.$ for $s\to\infty$. 
To proceed, we observe that the integral over $\xi$ is convergent since 
$\rho_1^-\sim\xi^{\alpha-1}$ and the expression in the curly bracket drops at 
least as $1/\xi^3$. Instead, the $x$ integral is peaked at 
$x\to0$, and we can therefore neglect $x$ in terms proportional to $(1-x)$. The divergent behavior of this integral  obtained in collinear approximation for the propagators 
  can obtained the formal  limit $Q^2\to\infty$  
   Then, the  expression in the curly bracket becomes $Q^2$-independent, 
and proportional to $\sim\ln(1-x_B)$  leading to a divergent integral of the type 
 $\int_0 dx x^{-\alpha}$.
To ensure convergence, we do not make this approximation. 
Changing finally the integration variable 
$x$ to $\omega=Q^2x/x_B$, we obtain Eq.(\ref{dvcs-f}).

\end{appendix}


\begin{thebibliography}{9}
\bibitem{footnote}
In the case of $a_2$, the subtraction is only necessary if the Pomeron has a 
non-zero magnetic coupling to the nucleon. Pomeranchuk's theorem rules
out Pomeron contribution to spin-flipping amplitudes.
The amplitude $a_4$ requires subtraction due to the $f_2$-exchange
with positive Regge intercept.  
Finally, $a_6$ corresponds to $\pi$-trajectory exchange with intercept 
$\alpha_\pi(0)\approx0$ and needs a subtraction. The amplitude $a_5$ 
is crossing-odd and needs no
subtraction since the corresponding Dirac structure does not allow for
a Pomeron contribution. However, a two-Pomeron exchange could
contribute to this amplitude.
\end{thebibliography}
\end{document}